\documentclass[sn-mathphys]{sn-jnl}% Math and Physical Sciences Reference Style
\jyear{2022}%

\theoremstyle{thmstyleone}%
\theoremstyle{thmstyletwo}%

\theoremstyle{thmstylethree}%

\begin{document}

\title[Strong enhancement of electromagnetic shower development induced by high-energy photons in a thick oriented tungsten crystal]{Strong enhancement of electromagnetic shower development induced by high-energy photons in a thick oriented tungsten crystal}

\author[1,2]{M. Soldani}
\author*[1]{L. Bandiera}\email{bandiera@fe.infn.it}
\author*[3]{M. Moulson}\email{matthew.moulson@lnf.infn.it}
\author[4,5]{G. Ballerini}
\author[6]{V.G. Baryshevsky}
\author[4,5]{L. Bomben}
\author[7,5]{C. Brizzolari}
\author[8]{N. Charitonidis}
\author[8]{G.L. D'Alessandro}
\author[9,10]{D. De Salvador}
\author[8]{M. van Dijk}
\author[3,11]{G. Georgiev}
\author[1]{A. Gianoli}
\author[1,2]{V. Guidi}
\author[6]{V. Haurylavets}
\author[6]{A.S. Lobko}
\author[12,13,1]{T. Maiolino}
\author[4,5]{V. Mascagna}
\author[1]{A. Mazzolari}
\author[1,2]{F.C. Petrucci}
\author[4,5]{M. Prest}
\author[1,14]{M. Romagnoni}
\author[15]{P. Rubin}
\author[16,17]{D. Soldi}
\author[1]{A. Sytov}
\author[6]{V.V. Tikhomirov}
\author[18]{E. Vallazza}

\affil[1]{Istituto Nazionale di Fisica Nucleare, Sezione di Ferrara, Ferrara, Italy}
\affil[2]{Dipartimento di Fisica e Scienze della Terra, Universit\`a degli Studi di Ferrara, Ferrara, Italy}
\affil[3]{Istituto Nazionale di Fisica Nucleare, Laboratori Nazionali di Frascati, Frascati, Italy}
\affil[4]{Istituto Nazionale di Fisica Nucleare, Sezione di Milano Bicocca, Milan, Italy}
\affil[5]{Dipartimento di Scienza e Alta Tecnologia, Universit\`a degli Studi dell'Insubria, Como, Italy}
\affil[6]{Institute for Nuclear Problems, Belarusian State University, Minsk, Belarus}
\affil[7]{Istituto Nazionale di Fisica Nucleare, Sezione di Milano Bicocca, Milan, Italy}
\affil[8]{CERN, Geneva, Switzerland}
\affil[9]{Istituto Nazionale di Fisica Nucleare, Laboratori Nazionali di Legnaro, Legnaro, Italy}
\affil[10]{Dipartimento di Fisica e Astronomia, Universit\`a degli Studi di Padova, Padua, Italy}
\affil[11]{Faculty of Physics, University of Sofia, Sofia, Bulgaria}
\affil[12]{School of Physics and Technology, Wuhan University, Wuhan, People’s Republic of China}
\affil[13]{WHU-NAOC Joint Center for Astronomy, Wuhan University, Wuhan, People’s Republic of China}
\affil[14]{Dipartimento di Fisica, Universit\`a degli Studi di Milano Statale, Milan, Italy}
\affil[15]{Department of Physics and Astronomy, George Mason University, Fairfax, Virginia, USA}
\affil[16]{Istituto Nazionale di Fisica Nucleare, Sezione di Torino, Turin, Italy}
\affil[17]{Dipartimento di Fisica, Universit\`a degli Studi di Torino, Turin, Italy}
\affil[18]{Istituto Nazionale di Fisica Nucleare, Sezione di Trieste, Trieste, Italy}

\abstract{We have observed a significant enhancement in the energy deposition by $25$--$100~\mathrm{GeV}$ photons in a $1~\mathrm{cm}$ thick tungsten crystal oriented along its  $\langle 111 \rangle$ lattice axes. At $100~\mathrm{GeV}$, this enhancement, with respect to the value observed without axial alignment, is more than twofold. This effect, together with the measured huge increase in secondary particle generation is ascribed to the acceleration of the electromagnetic shower development by the strong axial electric field. The experimental results have been critically compared with a newly developed Monte Carlo adapted for use with crystals of multi-$X_0$ thickness. The results presented in this paper may prove to be of significant interest for the development of high-performance photon absorbers and highly compact electromagnetic calorimeters and beam dumps for use at the energy and intensity frontiers.}




\maketitle

\section{Introduction}

High-$Z$ metals are widely used in high-energy physics for applications requiring their strength, radiation hardness and short radiation and nuclear interaction lengths. In particular, tungsten (W) is extensively exploited in a wide variety of applications: compact beam collimators \cite{Bruning15, Watanabe05, Shimogawa10}, radiation shielding \cite{Bruning15}, beam absorbers \cite{Watanabe05, Quaresima16} and targets for beam-dump experiments \cite{Ahdida19}. Moreover, tungsten foils with a thickness of one radiation length ($X_0 = 3.504~\mathrm{mm}$) or smaller are easy to manufacture and prove ideal as the passive absorber layers in sampling electromagnetic calorimeters, which are ubiquitous in high-energy physics experiments for the measurement of electron and photon energies. Similarly, thin tungsten foils have been used in space-borne $\gamma$-ray telescopes, such as in the AGILE \cite{Prest03} and FERMI LAT \cite{Atwood09} trackers, to convert incoming photons into ${e^+e^-}$ pairs. 

Currently, tungsten is employed as an amorphous material in all the aforementioned applications, while its crystalline nature is completely ignored. However, it has been well known since the 1950s that the electromagnetic interactions between high-energy particles and crystalline matter can be strongly affected by the atomic lattice structure of the latter \cite{TM}. The first coherent effect that was predicted theoretically \cite{TM} and then proved experimentally \cite{Palazzi68} is the so-called coherent bremsstrahlung (CB).  CB consists of the enhancement of the probability for bremsstrahlung emission that occurs when the $e^\pm$ is incident to the crystallographic planes or axes with a small angle $\theta$ and crosses different planes or axes with periodicity $d/\theta$, where $d$ is the interplanal or interaxial distance. Under these conditions, the momentum transferred by the ${e^{\pm}}$ to the crystal matches a reciprocal lattice vector, in analogy with Bragg-Laue diffraction. 
Similarly, for an incident photon under the same conditions, the pair production (PP) probability is enhanced, which is known as coherent pair production (CPP) \cite{Ferretti50}.

The CB (CPP) theories work as long as the nearly straight trajectory approximation is applicable to the motion of the incident (emitted) particle. However, this approximation fails \cite{Akhiezer96} when the charged particle trajectory is aligned with the crystal plane/axis within the so-called Lindhard critical angle \cite{lindhard65}, $\theta_\mathrm{L} = \sqrt{2 U_0 / E}$, with $U_0$ the depth of the potential well associated with the plane/axis and $E$ the projectile initial energy. Indeed, in this condition, the particle interacts coherently with the atoms in the same row/plane and is subject to transverse oscillations in the effective (averaged) electric planar/axial field of $\varepsilon \sim 10^{10}$--$10^{12}~\mbox{V}/\mbox{cm}$, i.e., channeling occurs \cite{BARYSHEVSKII1982153,PhysRevLett.50.950,Baryshevskii:1983JETP,Uggerhoj05}.

%On the other hand, when the particle velocity is nearly parallel, i.e., within the so-called Lindhard angle \cite{lindhard65}, to either an axis or a plane, the straight trajectory approximation, typical of CB, is no longer applicable: the charged particles are forced into an oscillatory motion within the planar or axial potential wells, i.e., channeling occurs.

%In fact, when the crystal is properly oriented, the Coulomb fields of the atoms in the plane or string add coherently. As a consequence, the $e^{\pm}$/$\gamma$ experience a continuous electric field that can reach the value of $\varepsilon \sim 10^{10}$--$10^{12}~\mbox{V}/\mbox{cm}$ \cite{BARYSHEVSKII1982153,PhysRevLett.50.950,Baryshevskii:1983JETP,Uggerhoj05}. The continuous field and the corresponding potential can also be used to describe the dynamics and radiation processes for unchanneled particles that enter the crystal with an angle larger than the one introduced by Lindhard.

\begin{figure}[htbp]
\centering
\includegraphics[width=\columnwidth]{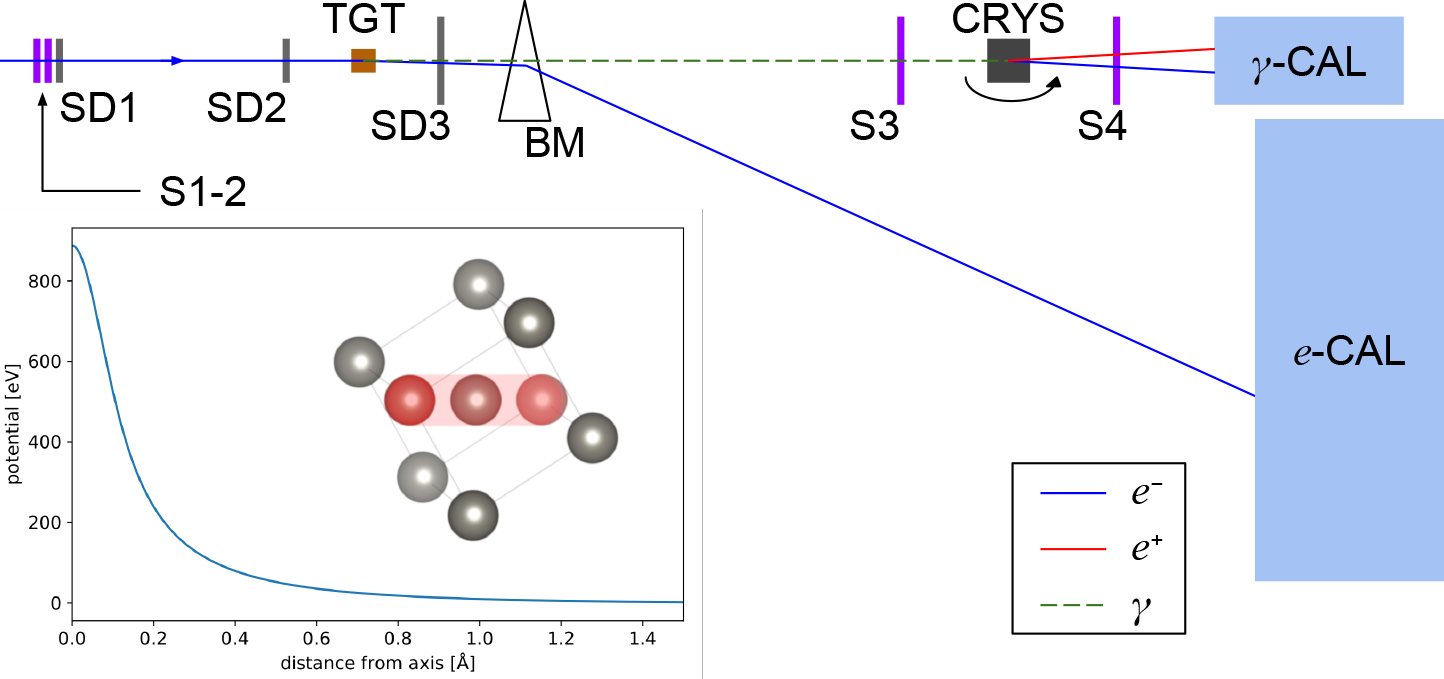}
\caption{Experimental setup at the CERN H2 beamline. The input electrons are incident on a thin copper radiator (TGT), producing bremsstrahlung photons; the latter are then separated from the charged beam by a bending magnet (BM). The S1--4 scintillating counters are used to generate the trigger signal (S1$\wedge$S2$\wedge \overline{\mbox{S3}}$) and to measure the downstream charged particle multiplicity (S4). The silicon microstrip detectors SD1--3 track charged particles. An array of lead-glass blocks measures the energy of the interaction products of the photon ($\gamma$-CAL) and of the deflected electron ($e$-CAL). Bottom left insertion: tungsten $\langle 111 \rangle$ axial potential as a function of the transverse distance from the axis and BCC lattice, with the $\langle 111 \rangle$ axis highlighted in red.}
\label{fig:SETUP}
\end{figure}

In the rest frame of the incident particle, this effective field is enhanced by a Lorentz factor, $\gamma_{\mathrm{eff}} = E / m c^2$ ($\gamma_{\mathrm{eff}} = \hbar \omega / m c^2$) for $e^\pm$ ($\gamma$), with $m$ the electron mass and $\omega$ the incident $\gamma$ frequency. Indeed, in QED, the pair production process in an external field is described as the conversion of the photon into a virtual electron-positron pair that subsequently becomes real through the interaction with the external field. Since the center of mass energy of the virtual pair amounts to $mc^2$, a Lorentz factor of about $E/mc^2$ is ascribed to the pair and can be used to describe the Schwinger limit for pair production as well.

At sufficiently high incident particle energy, the effective field can become comparable to the QED critical field, $\varepsilon_0 = m^2 c^3 / e \hbar \sim 1.32 \times 10^{16}~\mbox{V}/\mbox{cm}$, introduced by Schwinger \cite{Sauter31}. When $\gamma_{\mathrm{eff}}\varepsilon \geq \varepsilon_0 $, the strong crystal field gives rise to oscillatory motion of $e^+/e^-$ exceeding the opening angle $1/\gamma_{\mathrm{eff}}$ of the radiation cone, thereby leading to emission of synchrotron radiation, for which the quantum recoil of a radiating electron is not negligible \cite{Baryshevskii:1989, Akhiezer96, Baier98_book}. In practice, this occurs when the trajectory of the incident particle forms an angle $\theta$ with the lattice direction not much larger than $\Theta_0 = U_0 / mc^2$ \cite{BARYSHEVSKII1985430, BARYSHEVSKII1985335,BARYSHEVSKII1982697,Baier84}, where the axial field can indeed be considered constant over a distance larger than the typical length over which radiated photons are formed. In this strong field regime, there is a considerable enhancement in radiation emission, which exceeds both the standard Bethe-Heitler bremsstrahlung in amorphous media \cite{Bethe34} and the CB process in crystals. Due to the crossing symmetry of radiation emission and pair production, pair production is similarly enhanced for photons incident on the lattice at small angles to a main axis \cite{ Baier98_book,Uggerhoj05}.

The main features of the crystalline strong field can be described in terms of the parameter $\chi = \gamma_{\mathrm{eff}} \varepsilon / \varepsilon_0$ \cite{Ritus1985, Baier98_book}. For $\chi \ll 1$ the effective field is sub-critical, the synchrotron-like PP is strongly suppressed and the synchrotron-like radiation is soft (i.e., the emitted photon energy is a small fraction of the projectile energy). On the other hand, for $\chi \sim 1$ the critical field is reached and the situation changes dramatically: the synchrotron-like PP rapidly attains observable rates \cite{Baier84, Baier86} and the synchrotron-like radiation recoil becomes highly important, which results in an enhancement in the hard part of the photon energy spectrum. As a consequence, a strong acceleration of the electromagnetic shower development, or equivalently, a strong reduction of the radiation length, $X_0$, is attained. The resulting enhancement of the cross section grows with photon energies ranging from a few tens of GeV up to a saturation value in the multi-TeV region \cite{Baryshevskii:1989,KIMBALL198569,Uggerhoj05,Baier86,Akhiezer1980,Akhiezer1983}.

The enhancement of radiation and PP by high-energy $e^{\pm}$ and photons in strong and sub-strong crystalline fields was first experimentally investigated in the 1980s. The studies on PP were firstly focused on lighter elements such as silicon and germanium \cite{Belkacem87}, and only at a later stage on high-$Z$ crystalline metals such as tungsten and iridium \cite{Moore96, Kirsebom98, Baskov93}. Most of these studies exploited the axial field of the crystals, which is stronger than the planar field and therefore allows the strong field regime to be reached at relatively low energies. These studies were driven by the need to test the feasibility of an intense positron source \cite{CHEHAB200241, BOCHEK2001121} and to develop a compact photon converter, i.e., a device to separate the photon and neutral hadron beam components with minimal absorption or scattering of the latter \cite{Moore96, Kirsebom98}, for the NA48 experiment at CERN; eventually, an iridium crystal with a thickness of 0.98 $X_0$ was employed \cite{Fanti07}. Indeed, with the aim of demonstrating the radiation or PP enhancement, all these measurements featured samples with a thickness of $\lesssim 1 X_0$ or, in the case of a few studies limited to sub-strong field energies, slightly thicker ones, thus precluding a full investigation of the electromagnetic shower acceleration caused by the strong crystalline field and only allowing for either the radiative or the PP enhancement to be studied individually.

In this letter, we present an experimental investigation into the acceleration of the development of the electromagnetic shower occurring in interactions between $25$--$100~\mathrm{GeV}$ photons with a thick ($2.85~X_0$) tungsten crystal oriented along the $\langle 111 \rangle$ axes, i.e. in the full strong field regime --- $\chi \gtrsim 1$. The results were obtained at the CERN SPS, and demonstrate the strong enhancement of the electromagnetic shower development and, in particular, of the absorption power when the photon beam is aligned with the crystal axes. This effect might be advantageously exploited in future experiments at the energy and intensity frontiers, with applications in photon absorbers for fixed-target and beam dump experiments and in new generation detectors such as ultra-compact sampling calorimeters, as explained later in the text.

\section{Experimental setup}

Crystalline tungsten has a body-centered cubic (BCC) lattice (Fig.~\ref{fig:SETUP}, bottom left) with constant $a = 3.1652~\mbox{\AA}$. For our experiment we selected the $\langle 111 \rangle$ axis, which gives the strongest field in the case of W. The continuous potential associated with the $\langle 111 \rangle$ axis at room temperature is $U_0 = 887~\mbox{eV}$, which can be obtained from the curve plotted in Fig.~\ref{fig:SETUP}, bottom left; this corresponds to $\Theta_0 \sim 1.75~\mbox{mrad}$. For this axis, $\chi = 1$ at about $13.6~\mathrm{GeV}$ \cite{Baryshevskii:1989} at the maximum axial field $\varepsilon = \varepsilon_\mathrm{max} \sim 5 \times 10^{11}~\mbox{V}/\mbox{cm}$, and at $E_{\gamma} \gtrsim 22~\mathrm{GeV}$ the coherent PP probability becomes larger than the Bethe-Heitler value.

The commercial sample under study was produced by Princeton Scientific; it was approximately cubic, $10~\mbox{mm}$ (i.e., $2.85~X_0$) thick. The $\langle 111 \rangle$ axes were normal to one of the faces. It was tested at the CERN H2 beamline with a tagged-photon beam obtained from a tertiary $120~\mbox{GeV}/c$ electron beam. The experimental setup is shown in Fig.~\ref{fig:SETUP}.
To generate the tagged-photon beam, the electron beam was directed onto a 1-mm copper radiator.
The electrons were then magnetically deflected towards an array of lead-glass blocks \cite{Ereditato92}, which measured the electron energy after bremsstrahlung emission and therefore served as a photon tagging system ($e$-CAL in Fig. 1). 
Furthermore, the trajectories of the input electrons were reconstructed via two 
$\sim$ $20\times20$~mm$^2$ silicon microstrip sensors (SD1–2) with an overall angular resolution of a few $\mu$rad \cite{Lietti13}; this allowed the reconstruction of the impact point and trajectory of the photons incident on the crystal sample, given the small aperture of the bremsstrahlung cone from the copper target at this energy ($\sim$4 $\mu$rad).

The sample was mounted on a high-precision goniometer, which allowed positioning and orientation in both the horizontal and vertical planes to be controlled remotely with a resolution of $\lesssim 5~\mu\mbox{m}$ and $\lesssim 5~\mu\mbox{rad}$ respectively \cite{Lietti12, Bandiera13, Bandiera18}. Plastic scintillators were installed upstream and downstream with respect to the crystal position: the former (S3) served as a veto for the pairs produced by photon conversion in air (whose contribution was strongly suppressed by the use of vacuum pipes and a helium bag for transport of the deflected beam) whereas the latter (S4) allowed a measurement of the number of charged particles produced inside the crystal. Another lead-glass block ($\gamma$-CAL), placed along the photon beam trajectory, measured the energy transmitted in the forward direction.

The input beam divergence of $\sim$85 $\mu$rad ($\sim$60 $\mu$rad) in the horizontal (vertical) plane was considerably smaller than $\Theta_0$, allowing the dependence of the strong field on the input angle to be studied with high precision. The crystal was tested at several different values of the angle of incidence $\theta$ of the photon beam with respect to the $\langle 111 \rangle$ axis, ranging from 0 to 12.5 mrad ($\sim$7$\Theta_0$). Furthermore, data were collected at $\sim$45~mrad ($\sim$2.6$^\circ$) from the axis along a direction chosen to avoid proximity to other main axes and planes, where the lattice structure has no effect on the photon-matter interaction; this is equivalent to the case of random orientation.

\section{Results and simulations}

\begin{figure}[htbp]
%\includegraphics[width=\columnwidth]{figures/w111_calosCorr.pdf}
%\caption{Energy measured by $\gamma$-CAL as a function of the tagged photon energy, at different angles between the beam and the $\langle 111 \rangle$ axis and in the random orientation ($45~\mbox{mrad}$). The points (dashed lines) represent the experimental data (simulated results).}
\includegraphics[width=\columnwidth]{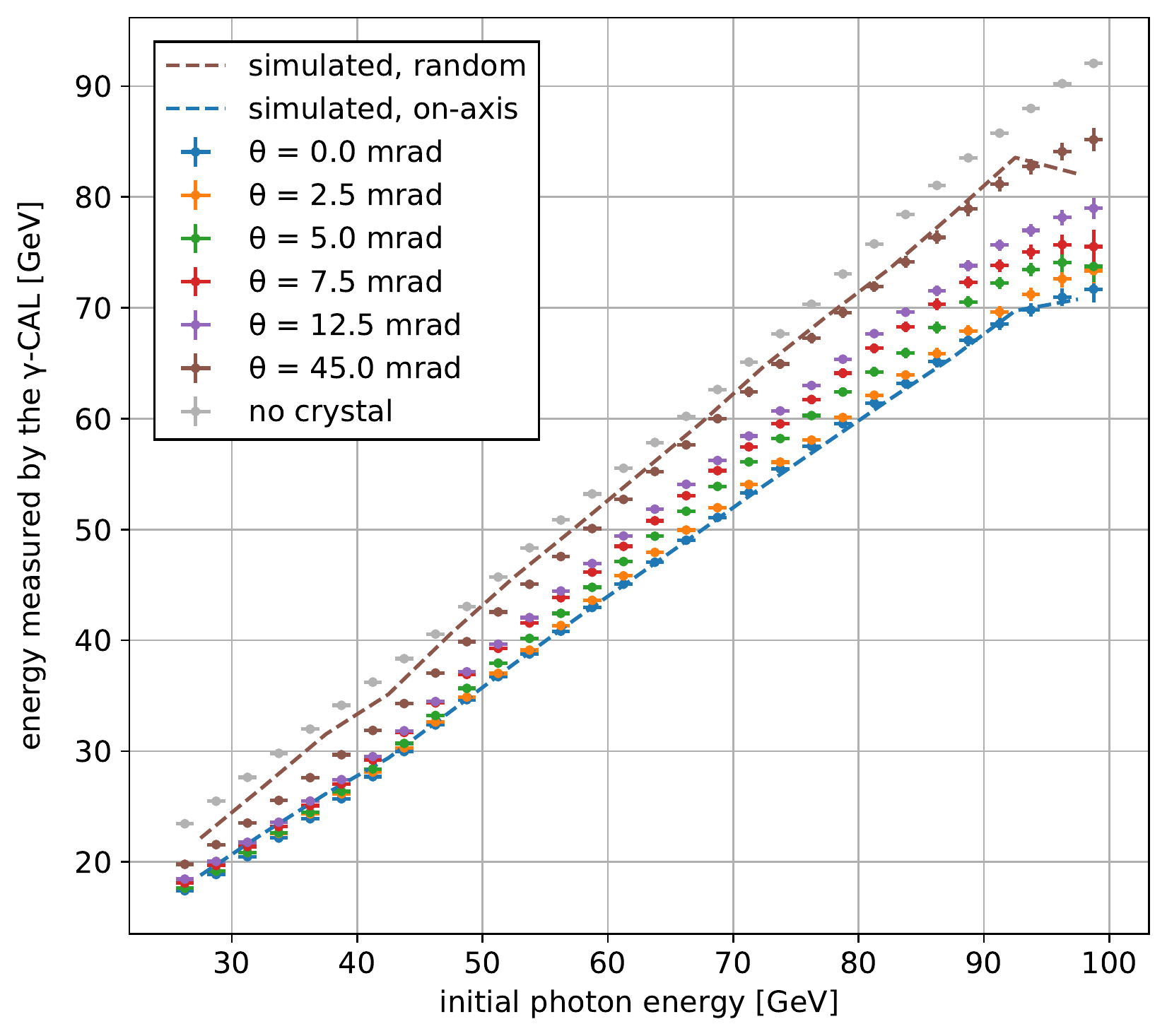}
\caption{Energy measured by $\gamma$-CAL as a function of the tagged photon energy, at different angles between the beam and the $\langle 111 \rangle$ axis and in the random orientation ($45~\mbox{mrad}$). Background data, i.e., with no crystal along the photon path, are also shown. The points (dashed lines) represent the experimental data (simulated results).}
\label{fig:PLOTS_CORR_A}
\end{figure}

Fig.~\ref{fig:PLOTS_CORR_A} shows the correlations between the transmitted energy directly measured by the $\gamma$-CAL and the initial photon energy. The latter quantity is obtained via photon tagging, i.e., by subtracting the electron energy measured in the $e$-CAL from the nominal value of the initial electron beam energy ($120~\mathrm{GeV}$). The energy absorbed or dispersed by the crystal, which we refer to as the lost energy, is given by $E_{\rm lost} = E_\gamma - E_{\rm bkg} - E_{\gamma\mathrm{-CAL}}$, where $E_\gamma$ is the initial photon energy from the photon tagging, $E_{\gamma{\rm-CAL}}$ is the transmitted energy as measured by the $\gamma$-CAL, and $E_{\rm bkg}$ is the missing energy $E_\gamma – E_{\gamma{\rm-CAL}}$ observed with no crystal sample in the apparatus. The lost energy is strongly enhanced when the beam is incident along the crystal axis (blue points), compared to the random orientation (brown points). To quantify, Table~\ref{tab:PLOTS_CORR_A} shows values of the lost energy (before background subtraction, with the background contribution separately tabulated) for three different values of the incident photon energy ($\sim$39~GeV, $\sim$69~GeV, and $\sim$94~GeV) and for different angles of incidence.
With the background subtracted, for a $\sim$94~GeV photon incident at $\theta = 0~\mathrm{mrad}$, the lost energy is about $18~\mathrm{GeV}$, while for random incidence, it is about $5~\mathrm{GeV}$. The amount of energy absorbed or dispersed by the crystal is seen to increase by about a factor of three. As clearly visible in Fig.~\ref{fig:PLOTS_CORR_A}, the maximum absorption power is maintained up to about $2.5~\mathrm{mrad}$ from the crystal axis, while the size of the effect decreases as the angle grows. Nevertheless, at $\sim$94~GeV, even at an angle of incidence of $12.5~\mathrm{mrad}$ (purple curve), the fraction of missing energy is still $1.5$ times that for the random orientation. As expected, with initial photon energy, since $X_0$ decreases, the amount of energy absorbed inside the crystal increases, so that the fraction of the initial photon energy absorbed is nearly constant over the whole range in $E_\gamma$ explored. The broad angular range over which the macroscopic character of the enhancement in energy absorption is preserved can be ascribed to both the strength of the $\langle 111 \rangle$ axis potential in tungsten and the high mosaicity of the sample, for which a value of $\sim$3~mrad was estimated from the combination of high-resolution measurements of X-ray diffraction on the crystal surface and Monte Carlo simulation. The slight saturation of the enhancement for $E_\gamma > 90~\mathrm{GeV}$ is due to the limited acceptance of the $e$-CAL, as we have verified via simulation.

\begin{table}[htbp]
\centering
\resizebox{\columnwidth}{!}{%
\begin{tabular}{r|ccc}
 & \multicolumn{3}{c}{missing energy in $\gamma$-CAL [GeV]} \\ \cline{2-4}
$\theta$ {[}mrad{]} & \multicolumn{1}{c|}{at $38.75~\mathrm{GeV}$} & \multicolumn{1}{c|}{at $68.75~\mathrm{GeV}$} & \multicolumn{1}{c}{at $93.75~\mathrm{GeV}$} \\ \hline
sim., on axis & \multicolumn{1}{c|}{$11.78$} & \multicolumn{1}{c|}{$17.78$} & \multicolumn{1}{c}{$23.76$} \\
exp., on axis & \multicolumn{1}{c|}{$13.05~(0.75)$} & \multicolumn{1}{c|}{$17.66~(0.81)$} & \multicolumn{1}{c}{$23.93~(0.96)$} \\
$2.5$ & \multicolumn{1}{c|}{$12.60~(0.75)$} & \multicolumn{1}{c|}{$16.78~(0.80)$} & \multicolumn{1}{c}{$22.54~(0.93)$} \\
$5.0$ & \multicolumn{1}{c|}{$12.38~(0.75)$} & \multicolumn{1}{c|}{$14.86~(0.80)$} & \multicolumn{1}{c}{$20.28~(0.94)$} \\
$7.5$ & \multicolumn{1}{c|}{$11.72~(0.76)$} & \multicolumn{1}{c|}{$13.44~(0.80)$} & \multicolumn{1}{c}{$18.72~(0.98)$} \\
$12.5$ & \multicolumn{1}{c|}{$11.32~(0.76)$} & \multicolumn{1}{c|}{$12.53~(0.80)$} & \multicolumn{1}{c}{$16.76~(0.91)$} \\
exp., random & \multicolumn{1}{c|}{$9.07~(0.80)$} & \multicolumn{1}{c|}{$8.75~(0.84)$} & \multicolumn{1}{c}{$11.00~(1.00)$} \\
sim., random & \multicolumn{1}{c|}{$6.32$} & \multicolumn{1}{c|}{$7.79$} & \multicolumn{1}{c}{$10.56$} \\
exp., no crystal & \multicolumn{1}{c|}{$4.62~(0.75)$} & \multicolumn{1}{c|}{$6.12~(0.76)$} & \multicolumn{1}{c}{$5.76~(0.79)$}
\end{tabular}
}
\caption{Missing energy (i.e., part of the initial photon energy not detected by the $\gamma$-CAL) at different angles between the beam and the $\langle 111 \rangle$ axis, for different values of the tagged photon energy. Errors are reported in parentheses.}
\label{tab:PLOTS_CORR_A}
\end{table}

The experimental results presented in Fig.~\ref{fig:PLOTS_CORR_A} demonstrate the faster electromagnetic shower development due to the strong crystalline field, with a resulting enhancement of secondary particle generation in the first layer of material. Each of these charged secondaries deposits energy inside the material during its passage, resulting in a significant increase of absorbed energy when the sample is oriented along its $\langle 111 \rangle$ axes.

\begin{figure}[htbp]
\includegraphics[width=\columnwidth]{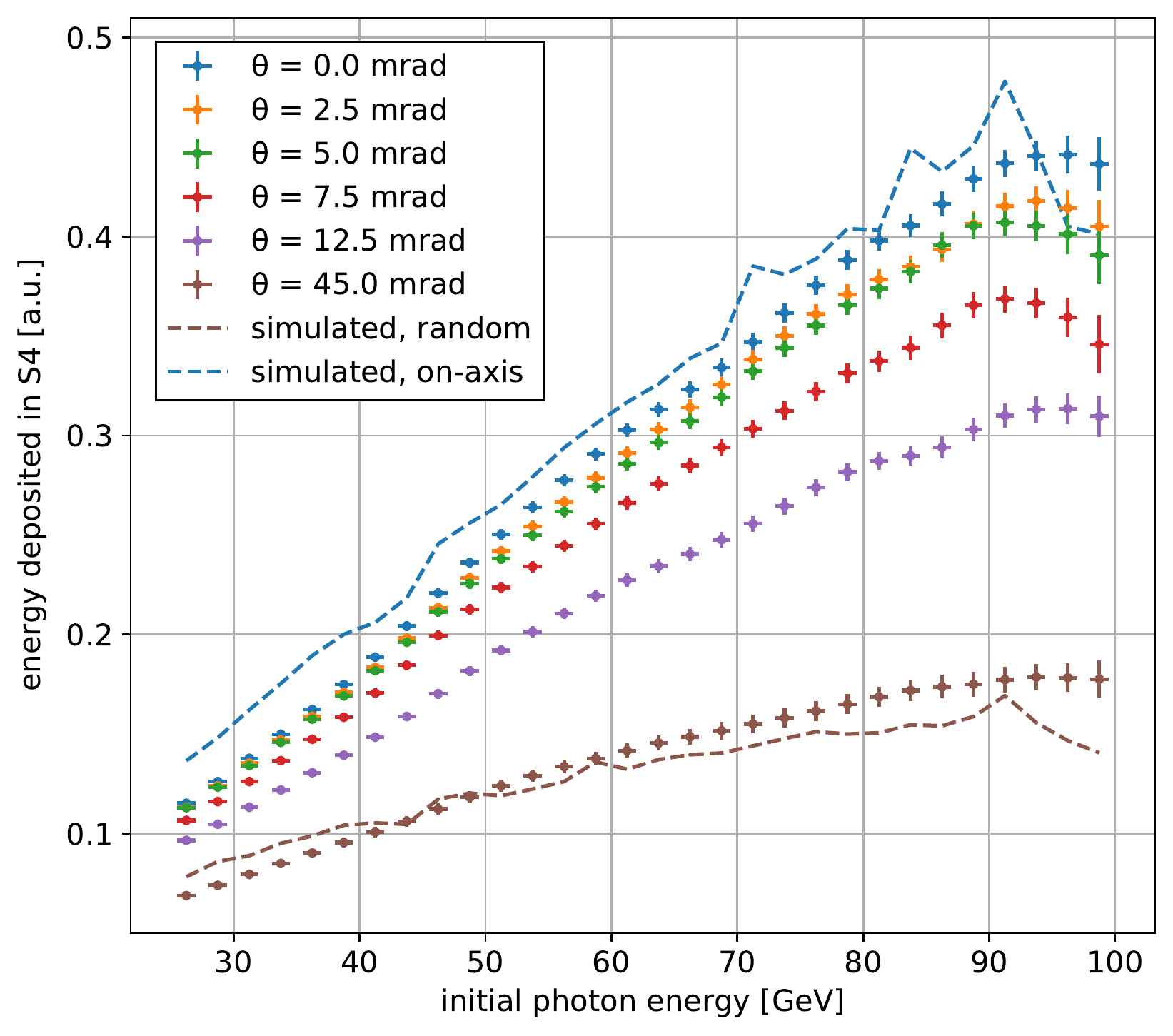}
\caption{Energy deposited in the S4 plastic scintillator as a function of the tagged photon energy, at different angles between the beam and the $\langle 111 \rangle$ axis and in the random orientation ($45~\mbox{mrad}$). The points (dashed lines) represent the experimental data (simulated results).}
\label{fig:PLOTS_CORR_B}
\end{figure}

We also directly measured the increase in the secondary production due to the acceleration of shower development with the S4 scintillating multiplicity counter, whose pulse height is proportional to the energy deposited inside the plastic layer and therefore to the number of incident charged particles (Fig.~\ref{fig:PLOTS_CORR_B}). As expected, the enhancement grows with the tagged photon energy, i.e., at higher initial $\chi$, and for decreasing angle of incidence with respect to the axis. The corresponding ratios between measured values at different angles of incidence and in the random orientation (i.e., $45~\mathrm{mrad}$ off axis) range from $130$--$160\%$ at $\sim$30~GeV, depending on the incoming photon angle, to $\gtrsim\,$230\% at $100~\mbox{GeV}$ when on axis; indeed,  even at an angle of $12.5~\mathrm{mrad}$ with respect to the axis, the enhancement ratio is $\sim$170\% for $100~\mathrm{GeV}$ photons. Again, the apparent saturation of the enhancement for $E_\gamma > 90~\mathrm{GeV}$ is due to the limited acceptance of the $e$-CAL.

All of the experimental results presented here are validated by the agreement with the results of Monte Carlo simulations (see dashed lines in Figs.~\ref{fig:PLOTS_CORR_A} and \ref{fig:PLOTS_CORR_B}). The simulation of the full experimental setup was based on the Geant4 toolkit \cite{Geant03}, in which the features of the radiation-matter interaction for amorphous media is implemented by default. The results for the axial configuration were obtained by rescaling the bremsstrahlung and PP cross sections according to the outcome of full Monte Carlo computations based on the Baier-Katkov quasi-classical operator method \cite{Baier84,Baier86} to simulate the enhancement in oriented crystals; see, e.g., \cite{Baryshevskii17, Bandiera19}. Indeed, since the crystal mosaicity exceeded the angle $\Theta_0$, our experimental conditions were far from those for which the uniform field approximation is applicable \cite{Uggerhoj05}. For this reason, the full scale evaluation of the Baier-Katkov formulae was used in the simulations. 
Given the considerable mosaicity of the thick crystal under study in this work, both the radiation and pair production probabilities were averaged over the distribution representing the spread of the angle of incidence arising from the crystal mosaic and implemented in Geant4 instead of the Bethe-Heitler values. Geant4 proved essential in combining the coherent radiation and pair production enhancements at the high particle energies present at the start of the shower with the highly efficient and well-verified treatment of the energy loss for the numerous sub-GeV particles that mostly move at large angles with respect to the trajectory of incidence in the advanced stages of the shower.

\section{Outlook for applications}

The observation of the enhancement in photon energy absorption and secondary pair production described in this work paves the way to manifold applications in high-energy physics and astrophysics. While the strong field effects result in a reduction of the effective $X_0$ that depends on the particle energy, the nuclear interaction length $\lambda_{\mathrm{int}}$, i.e., the longitudinal scale of the inelastic hadronic processes, is unaffected by the lattice orientation. Therefore, axially oriented tungsten layers prove appealing to filter out the photon component in hadron beams with minimum effect on the hadronic component \cite{Ambrosino19}. As an example, this technique could be used to clean the prompt photons from a neutral hadron beam as proposed by the KLEVER ($K_L$ Experiment for VEry Rare events) experiment, which is planned to follow upon the NA62 experiment in the CERN North Area, extending the program of searches for rare kaon decays to include $K_L$ mesons. \cite{Ambrosino19}.

Similarly, beam-dump experiments for light dark matter searches could benefit from an oriented crystal dump target, which would allow for an increase in the experiment sensitivity; indeed, a crystalline target would allow the dump length to be reduced without any reduction in its electromagnetic shower containment capability and therefore the probability to be increased for dark photons to exit the dump before decaying, so that their decay products can be reconstructed in the apparatus downstream \cite{Nardi18}. 
Moreover, it is well known that, at high-energies, bremsstrahlung and PP are suppressed in amorphous media due to the Landau-Pomeranchuk-Migdal (LPM) effect \cite{Landau53}. Since bremsstrahlung and PP are dramatically enhanced in crystals at high energy and at the same time are not reduced by the LPM effect, the use of aligned crystals is appealing for the
design of beam dumps for use at the energy frontier.

\begin{figure}[htbp]
\includegraphics[width=\columnwidth]{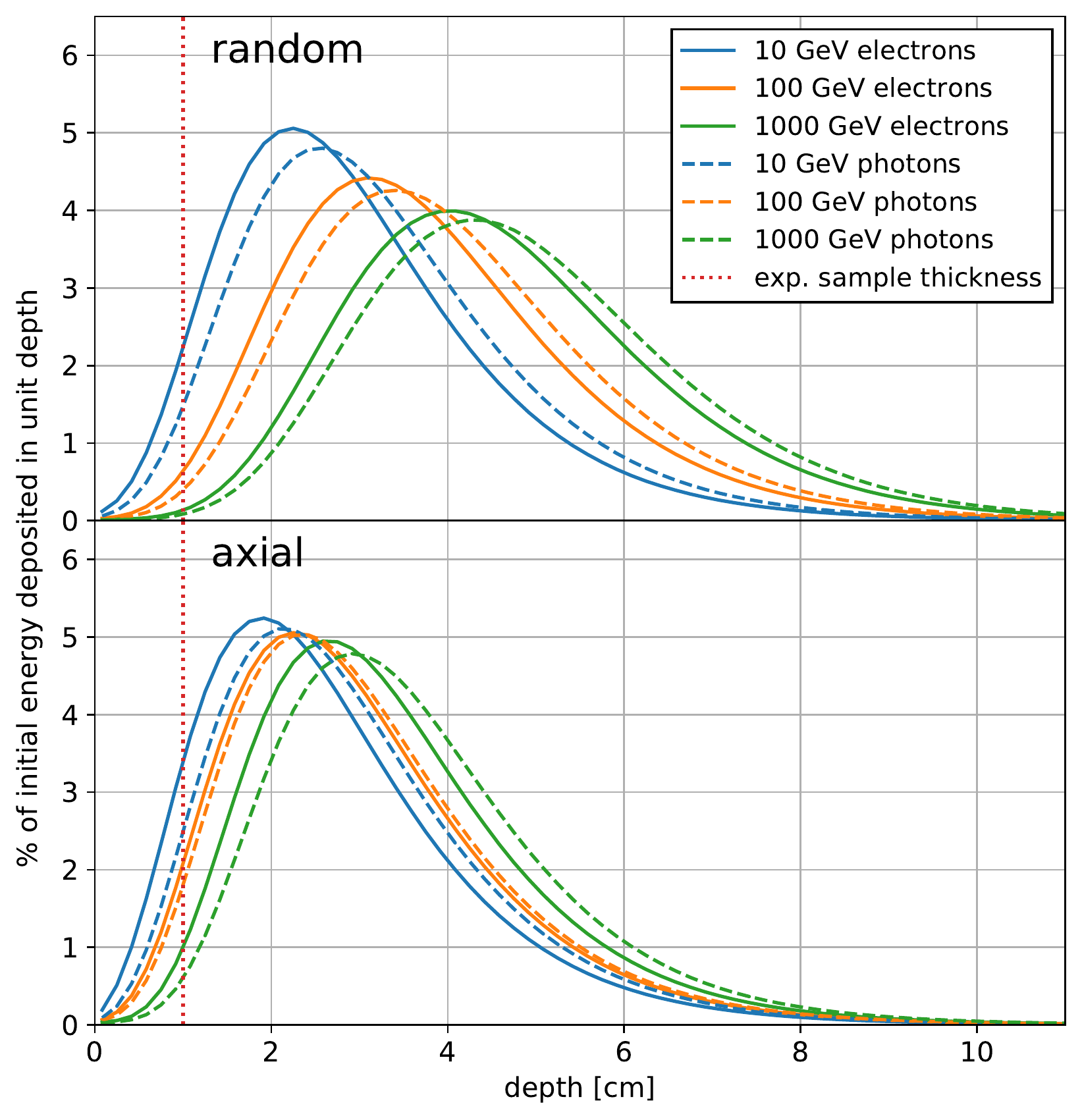}
\caption{Simulated energy absorbed in crystalline tungsten oriented along the $\langle 111 \rangle$ axis, normalized to the primary particle energy and to the unit depth, as a function of the depth inside the crystal, for incident electrons (solid curves) and photons (dashed curves). The simulated crystal parameters are the same as in the experimental case presented in Figs.~\ref{fig:PLOTS_CORR_A} and \ref{fig:PLOTS_CORR_B}.}
\label{fig:SHOWERS}
\end{figure}

The reduction in electromagnetic shower length is clearly visible in Fig.~\ref{fig:SHOWERS}, which shows the simulated energy deposited in crystalline tungsten oriented along the $\langle 111 \rangle$ axis by primary electrons/photons. The curves are obtained from simulation of highly collimated electron and photon beams incident on a $20 \times 20 \times 20~\mathrm{cm}^3$ tungsten crystal, i.e., wide enough to contain all the shower particles, oriented along the $\langle 111 \rangle$ axis: the fraction of the absorbed primary particle energy is plotted, at different input energies and for the random (top) and axial (bottom) cases, as a function of the depth inside the material. When on axis, the maxima of all of the curves are located approximately at the same depth, thus demonstrating that in case of an axially oriented crystal the position of the shower maximum is nearly independent of the initial energy between a few GeV and $\sim 1~\mathrm{TeV}$. For a $100~\mathrm{GeV}$ photon, the energy deposition at a depth of $1~\mathrm{cm}$ is about three times higher in the axial orientation than in the random orientation and corresponds to that of an amorphous target of length $\sim 1.47~\mathrm{cm}$. This enhancement is basically in good agreement with our observed value.
%, due to the limited transverse size of the sample tested and of the $\gamma$-CAL acceptance.
The same simulation code allows an estimate to be obtained of the effective radiation length from the fraction of incident $100~\mathrm{GeV}$ photons that cross $1~\mathrm{cm}$ of tungsten without converting into ${e^+e^-}$ pairs: for the axial case, a value of $1.050~\mathrm{mm}$ was obtained, i.e., $3.3$ times the amorphous value.

The acceleration of the shower development can be exploited to create sampling electromagnetic calorimeters with thinner passive layers. Sampling calorimeters are widely used in particle physics for the measurement of particle energies and consist of several high-$Z$, high-density passive layers, which force incoming electromagnetic particles to shower, interleaved with active layers (typically plastic scintillator) in which a signal proportional to the number of particles per stage is generated. The use of oriented tungsten absorber layers instead of amorphous ones would allow for a significant reduction of the calorimeter thickness, leading to the development of ultra-compact calorimeters with applications in fixed-target experiments, at the energy and intensity frontiers. Furthermore, as seen from Fig.~\ref{fig:SHOWERS}, in a sampling calorimeter using oriented crystal absorbers, the depth profile of the electromagnetic shower is nearly independent of the incident particle energy, thus reducing or eliminating non-linearities due to shower leakage for high-energy particles. The results presented in this letter demonstrate that metallic crystals that are already commercially available are suitable for this purpose. The possibility to assemble many crystals in a matrix and to control their reciprocal orientation with an accuracy better than $\sim 1~\mathrm{mrad}$ can be achieved with existing technologies such as bonding techniques \cite{Robertson}, thereby allowing realistic applications in large converter and detector systems.

\section{Conclusions}

In summary, we have presented measurements of the acceleration of electromagnetic shower development for $25$--$100~\mathrm{GeV}$ photons incident on a thick (multi-$X_0$), commercially available tungsten crystal at small angle relative to the $\langle 111 \rangle$ axis, demonstrating an enhancement in the secondary particle production and energy absorption for axially oriented crystals, compared to the case for random orientation. This behaviour is maintained over an angular acceptance of at least a few mrad. The magnitude of the observed effect is promising for the use of thick, commercial crystalline tungsten layers as compact photon or electron absorbers in current and future fixed-target experiments, which are intrinsically forward. These results also confirm that high-$Z$ crystals such as tungsten prove interesting for the construction of next-generation, ultra-compact sampling calorimeters and targets for beam-dump experiments.

\section*{Acknowledgements}

This work was partially supported by INFN CSN1 (NA62 experiment/KLEVER project) and CSN5 (STORM/OREO/MC-INFN projects), the CERN Physics Beyond Colliders initiative, the US NSF (grants no. 1506088, 1658621 and 1806430) and the European Commission through the H2020-MSCA-RISE N-LIGHT (G.A. 872196) and EIC-PATHFINDER-OPEN TECHNO-CLS (G.A. 101046458) projects. We acknowledge the support of the CERN PS/SPS physics coordinator, of the BE-OP-SPS physicists and operators and of the BE-EA group technical staff. M. Romagnoni acknowledges support from the ERC Consolidator Grant SELDOM G.A. 771642. A. Sytov acknowledges support from the H2020-MSCA-IF-Global G.A. 101032975. We acknowledge the CINECA award, under the ISCRA initiative, for the availability of high performance computing resources and support. We thank our IRES students K. Ayers and G. Quaresima for their participation in data taking, analysis and simulation.

\section*{Data Availability Statement}

The manuscript has associated data in a data repository. [Authors’ comment: \url{https://zenodo.org/record/7541067#.Y9jWyXbMJGN}.]

\section*{Open Access}

This article is licensed under a Creative Commons Attribution 4.0 International License, which permits use, sharing, adaptation,
distribution and reproduction in any medium or format, as long as you
give appropriate credit to the original author(s) and the source, provide a link to the Creative Commons licence, and indicate if changes
were made. The images or other third party material in this article
are included in the article’s Creative Commons licence, unless indicated otherwise in a credit line to the material. If material is not
included in the article’s Creative Commons licence and your intended
use is not permitted by statutory regulation or exceeds the permitted use, you will need to obtain permission directly from the copyright holder. To view a copy of this licence, visit \url{http://creativecommons.org/licenses/by/4.0/}.

Funded by SCOAP3. SCOAP3 supports the goals of the International
Year of Basic Sciences for Sustainable Development.

\bibliography{biblio}

\end{document}